\documentclass[prl,twocolumn,showpacs,preprintnumbers,amsmath,amssymb]
{revtex4}

\usepackage{graphicx}
\usepackage{epstopdf}
\usepackage{amsmath}
\usepackage{amsbsy}
\usepackage{dcolumn}
\usepackage{bm}

\begin{document}

\title{Disorder effects on the intrinsic nonlinear current density in YBa$_2$Cu$_3$O$_{7-\delta}$}

\author{Brian M. Andersen,$^1$ James C. Booth,$^2$ and P. J. Hirschfeld$^1$}
\affiliation{$^1$Department of Physics, University of Florida,
Gainesville, Florida 32611-8440, USA\\ $^2$National Institute of
Standards and Technology, Boulder, Colorado 80303, USA }

\date{\today}

\begin{abstract}
We present harmonic generation measurements of the intrinsic
nonlinear current density $j_2$ of YBa$_2$Cu$_3$O$_{7-\delta}$
(YBCO) films at temperatures close the $T_c$. Experiments on a
range of different quality samples allow us to extract the
dependence of $j_2$ on the penetration depth of the
superconductor. In order to model these results, we calculate the
intrinsic nonlinear current response of $d_{x^2-y^2}$-wave
superconductors in the Meissner regime in the presence of
nonmagnetic impurities within the self-consistent T-matrix
approximation.
\end{abstract}


\maketitle

\subsection{Introduction}

Superconductors with $d$-wave pairing symmetry should exhibit an
unusual angle dependent nonlinear Meissner
effect\cite{yipsauls,Xu,stojkovic}. This has the effect of
significantly altering the current density dependence of the
superfluid density at low temperatures, and dramatically affects
the nonlinear response of superconductor transmission lines or
resonators. Recently, intermodulation measurements of resonators
fabricated from epitaxial YBCO thin films revealed a sharp
increase in the observed nonlinearity at low
temperatures\cite{benz,oatesprl,boothasc}, a result that was
originally predicted by Dahm and
Scalapino\cite{DahmScalaPRB,DahmScalaJAP}. While the measured
temperature dependence suggests that this observed nonlinear
response has intrinsic origins, the effect of impurities on the
nonlinear behavior in high $T_c$ superconductors has not been
explored systematically by experiments. This issue is important
since the nonlinear response and its unwanted intermodulation
products can severely limit the applications of superconducting
devices used for e.g. communication filters.

In this paper we report on measurements of the nonlinear current
density scale $j_2$ determined from third-harmonic generation
experiments for a range of YBCO thin-film samples at a fixed
temperature (76 K), but which display varying levels of
growth-induced defects, reflected in the measured penetration
depth. The samples were fabricated under a variety of growth
conditions, and showed variations in the penetration depth over
the range from 0.25 - 0.4 $\mu$m at 76 K. Subsequent
third-harmonic generation measurements on these samples allowed us
to examine the nonlinear current density scale $j_2$ as a function
of the measured penetration depth, in order to explore the effects
of impurities on the nonlinear response.  We compare these
measured results with theoretical predictions for a
$d_{x^2-y^2}$-wave superconductor in the Meissner regime in the
presence of nonmagnetic impurities.

For a superconductor, the superfluid density is a function of the
current density. This manifests itself experimentally as a
dependence on current density of the superconductor magnetic
penetration depth.  The exact form of the current-dependent
penetration depth depends on the nature of the superconductor,
including the details of  the superconductor energy gap. Since for
most cases the current-density dependence of the penetration depth
$\lambda(j)$ is very small, we can use a polynomial approximation
for this quantity. We expect the change in the penetration depth
to be an even function of the current density $j$, so the
lowest-order term in this low-temperature expansion is quadratic:
\begin{equation}
\left( \frac{\lambda(j)}{\lambda_0}
\right)^2=1+(j/j_2)^2\label{lambdaexp},
\end{equation}
where $\lambda_0$ is the zero-current penetration depth and $j_2$
sets the scale of the nonlinear current density ($j_2$ is
sometimes also referred to as the intermodulation critical
current\cite{hammond}). In a $d$-wave superconductor,
Eq.(\ref{lambdaexp}) is known to break down at low temperatures
due to singular contributions to the current from nodal
quasiparticles. The nonlinear current density scale $j_2$
characterizes the strength of the current-density dependence of
the penetration depth, and its behavior as a function of e.g.
temperature and disorder can reveal important details regarding
the nature of the superconducting state. It is the measured
temperature dependence of $j_2$ extracted from nonlinear microwave
experiments that reveals the $d$-wave behavior predicted by Dahm
and Scalapino\cite{DahmScalaPRB,DahmScalaJAP}. In the present work
we examine the dependence of $j_2$ on the penetration depth in
order to explore the role of disorder in the nonlinear response of
high $T_c$ superconductors.

\subsection{Experiment}

Experimentally, there are a number of different ways to measure
$\lambda(j)$.  However, the large value for $j_2$ that results
from intrinsic effects\cite{DahmScalaJAP} means that very large
current densities are required to produce even very small changes
in $\lambda(j)$. This means that extrinsic effects due to
self-heating or vortex motion, for example, may easily mask the
intrinsic behavior, making the extraction of reliable values for
$j_2$ extremely challenging. One technique that possesses the
required sensitivity to small changes in $\lambda$ is the mutual
inductance technique reported by Claassen $\it et$ $\it
al.$\cite{Claassen1}, which used a persistent current trapped in
the thin-film samples to measure $\lambda(j)$ as a function of dc
current. By carefully accounting for heating effects of the dc
current induced in the measurement coils this technique was able
to extract an accurate value for $j_2$. The experiments
demonstrated the expected quadratic dependence of penetration
depth on dc current density\cite{Claassen1}.

It has also been realized that a current-density dependent
penetration depth will result in a current-dependent inductance
per unit length for a superconducting thin-film transmission
line\cite{DahmScalaJAP,Collado}, or a current-dependent film
inductance for mutual inductance experiments\cite{Claassen2}. Such
a nonlinear inductance will produce mixing or harmonic products
for single-frequency or two-tone stimuli, which can be detected
with remarkable sensitivity using straightforward spectrum
analysis\cite{BoothJAP}  or lock-in techniques.  Use of such
harmonic mixing or harmonic generation techniques means that
nonlinear effects generated by even a very small current-dependent
contribution to the penetration depth can be detected fairly
easily even for modest applied current densities.  In what
follows, $j_2$ determined from both third harmonic generation
experiments at microwave frequencies, as well as from mutual
inductance measurements at audio frequencies, is examined for YBCO
samples that contain differing amounts of growth-induced defects,
as reflected in the measured penetration depth.

\begin{figure}
\includegraphics[width=8cm,height=7cm]{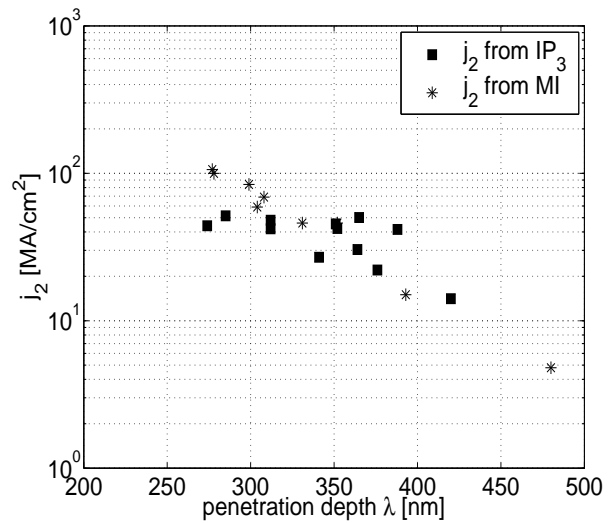}
\caption{Intrinsic nonlinear current density scale $j_2$ versus
penetration depth $\lambda$ extracted from a range of YBCO samples
of different quality. Here we show the results obtained from the
measured third harmonic signals $(IP_3)$ at 76 K and the mutual
inductance (MI) results by Claassen\cite{Claassen2} at 77
K.\label{j2vslambda}}
\end{figure}

For the microwave results described here, we use patterned
coplanar waveguide transmission lines of different lengths and
cross-sectional geometries fabricated from YBCO films with
thicknesses in the range 50-400 nm. The YBCO thin films were grown
by pulsed laser deposition, as well as by reactive co-evaporation.
All of the films were grown on LaAlO$_3$ substrates, and all had
superconducting transition temperatures in the 88-90 K range. Some
films were intentionally grown under different growth
conditions\cite{BoothASC98} while others displayed difference that
were simply the result of random process variations. For the
microwave transmission line measurements, we apply a single-tone
incident signal (typically 2-6 GHz), and measure the transmitted
signal at the fundamental frequency as well as at the third
harmonic frequency. From the measured third harmonic signal versus
incident power, along with the transmission line dimensions and
measured penetration depth, we can extract
$j_2$\cite{BoothOrigins}. For transmission lines of different
lengths and cross-sectional geometries fabricated from the same
thin-film sample, we obtained consistent values\cite{BoothJAP} for
$j_2$.  We also obtained consistent $j_2$ values from results of
two-tone intermodulation measurements in transmission-line
resonators, as well as from analysis of the power-dependent
impedance derived from resonator measurements performed as a
function of rf power\cite{BoothOrigins}.  The self-consistency of
the results for $j_2$ extracted from these many different
measurements suggests that we are measuring an intrinsic,
material-dependent nonlinear effect, as opposed to extrinsic
nonlinear effects due to e.g. vortex motion or heating effects.
The results of $j_2$ measurements at 76-77 K are shown for a
number of different samples in Fig. \ref{j2vslambda}, as a
function of the measured penetration depth (also determined at
76-77 K).

As mentioned above, mutual inductance measurements at audio
frequencies can also provide a measure of $j_2$\cite{Claassen2}.
For these measurements, the unpatterned superconducting film is
sandwiched between two coils driven at a frequency in the audio
range (typically 10 kHz).  Such an experimental arrangement can be
used to derive $j_2$ in two different ways: (1) by measuring
changes in the penetration depth as a function of a dc current
induced in the superconducting film, or (2) by measuring the
signal generated at the third harmonic of the drive signal using a
lock-in amplifier. The two approaches have been
shown\cite{Claassen2} to give consistent results for $j_2$ at 76
K, and $j_2$ measured by the former procedure has been shown to
agree quantitatively with the third-harmonic generation results at
microwave frequencies\cite{Claassen1}. The results of the mutual
inductance determination of $j_2$ for a variety of samples from
Ref. \cite{Claassen2} are also plotted in Fig. \ref{j2vslambda} as
a function of the measured penetration depth. The agreement
demonstrated between the microwave-frequency harmonic generation
measurements and the mutual inductance techniques for $j_2$ values
further supports our assertion that the nonlinear effects
quantified by $j_2$ are intrinsic in nature. For the films
measured by the mutual inductance technique, the penetration depth
was measured also by mutual inductance\cite{Claassen_lambda}. For
the films measured by the microwave third harmonic technique, the
penetration depth was measured by mutual inductance and by a
microwave transmission line technique\cite{Booth_lambda}. For both
cases, the penetration depth was measured at sufficiently low
signal levels to ensure that only linear contributions to the
penetration depth were recorded. In Fig. \ref{j2vslambda} it is
evident that, independent of the applied measurement technique,
there is a clear trend showing that films with smaller penetration
depths yield greater values for $j_2$.  In what follows we show
that these results can be modelled within conventional $d$-wave
BCS theory by including the role of defects.

\subsection{Theory}

The total current density can be written
as\cite{Xu,stojkovic,DahmScalaPRB}
\begin{eqnarray}\label{current}\nonumber
j=&-&2j_c \int_{-\pi/2}^{\pi/2} \frac{d \Theta}{2\pi} \cos \Theta
\int_{-\infty}^{\infty} \frac{d
\omega}{\Delta_0} f (\omega) \left[ N_+(\Theta,\omega) \right.\\
&-& \left. N_-(\Theta,\omega)\right],
\end{eqnarray}
where $f(\omega)$ is the Fermi function and
$N_{\pm}(\Theta,\omega)$ denote the density of states for comoving
and countermoving quasiparticles:
\begin{equation}
N_{\pm}(\Theta,\omega)=\mbox{Im}\frac{\tilde{\omega} \pm
\Delta_0(j_s/j_c)\cos \Theta}{\sqrt{\Delta^2(\Theta)-\left[
\tilde{\omega} \pm \Delta_0(j_s/j_c)\cos \Theta \right]^2}}.
\end{equation}
Here, $j_s$ is the superfluid current density and $j_c$ the
thermodynamic critical current density. In the following we will
be interested in the current response of dirty superconductors and
perform the calculations on the imaginary axis. Then the
associated Matsubara frequencies $\tilde{\omega}_n$ are
renormalized and must be determined from the following
selfconsistency conditions\cite{DahmScalaPRB}:
\begin{eqnarray}\label{renormmat}
\tilde{\omega}_n&=&\omega_n+\Gamma
\frac{g_0}{c^2+g_0^2},\\\label{renormgreen} g_0&=&\int_{-\pi}^\pi
\frac{d \Theta}{2\pi} \frac{\tilde{\omega}_n
}{\sqrt{\Delta^2(\Theta) + \tilde{\omega}_n^2}},
\end{eqnarray}
where $\omega_n=(2n+1)\pi T$ denote the bare fermion Matsubara
frequencies. As usual, $\Gamma=n_i/(\pi N_0)$ is the scattering
rate proportional to the impurity concentration $n_i$, and $c=\cot
\delta_0$, where $\delta_0$ is the s-wave scattering phase shift
and $N_0$ is the density of states at the Fermi level. The strong
scattering limit (unitary limit) is characterized by $c=0$,
whereas for the Born limit $c \gg 1$. For simplicity we have
assumed a circular Fermi surface and study current flow along the
nodal direction. The d$_{x^2-y^2}$-wave order parameter relevant
for the high-T$_c$ cuprate materials can be written as
\begin{equation}
\Delta(\Theta)=\Delta_0 \sin (2 \Theta).
\end{equation}

We have included the suppression of $T_c$ and $\Delta_0$ by
solving Eq. (\ref{renormmat}) and (\ref{renormgreen}) together
with the self-consistent gap
equation\cite{hotta,fehrenbacher,maki,bma}
\begin{equation}\label{gapeqn}
\frac{1}{g}=2\pi T\sum_n \int_0^{2\pi} \frac{d\Theta}{2\pi}
\frac{\sin^2 (2\Theta)}{\sqrt{\tilde{\omega}_n^2+\Delta^2
(\Theta)}},
\end{equation}
where $g$ denotes the coupling constant and the sum is restricted
by an upper energy cut-off $\omega_c$.

The superfluid density $n_s(j_s,T)$ is defined from the
relation\cite{DahmScalaPRB,DahmScalaJAP}
\begin{equation}
j = \frac{n_s(j_s,T)}{n} j_s,
\end{equation}
and in the local approximation, is related to the London
penetration depth $\lambda(j_s,T)$ by
\begin{equation}
\left( \frac{\lambda(j_s,T)}{\lambda_0} \right)^2 =
\frac{n}{n_s(j_s,T)},\label{lambdadef}
\end{equation}
where $n$ is the total electron density and $\lambda_0$ denotes
the zero-current penetration depth. In agreement with Eq.
(\ref{lambdaexp}) and (\ref{lambdadef}), it is through the
expansion of the superfluid density $n_s$ in terms of $j_s$ that
$j_2$ is defined by the following relation
\begin{equation}
n_s(j_s,T) = n_s(T) \left[1 - \left( \frac{j}{j_2} \right)^2
\right].\label{nsexp}
\end{equation}
Here, $j_2$ can be expressed as
\begin{equation}\label{j2def}
\left(\frac{j_2}{j_c} \right)^2=\frac{(n_s(T)/n)^3}{\beta(T)},
\end{equation}
where
\begin{equation}
\frac{n_s(T)}{n}=2 \pi \int_{-\pi}^\pi \!\! \frac{d \Theta}{2\pi}
T \sum_n \mbox{Re} \left[ \frac{\Delta^2 (\Theta)\cos^2
\Theta}{\left(\Delta^2(\Theta) + \tilde{\omega}_n^2 \right)^{3/2}}
\right],
\end{equation}
and
\begin{eqnarray}
\beta(T)&=&2 \pi \Delta_0^2 \int_{-\pi}^\pi \!\! \frac{d
\Theta}{2\pi} \Delta^2 (\Theta) \cos^4 \Theta \label{betadef}\\
\nonumber &\times& T \sum_n \mbox{Re} \left[
\frac{4\tilde{\omega}_n^2 - \Delta^2
(\Theta)}{\left(\Delta^2(\Theta) + \tilde{\omega}_n^2
\right)^{7/2}} \right].
\end{eqnarray}
\begin{figure}[t]
\includegraphics[width=8.0cm,height=6.0cm]{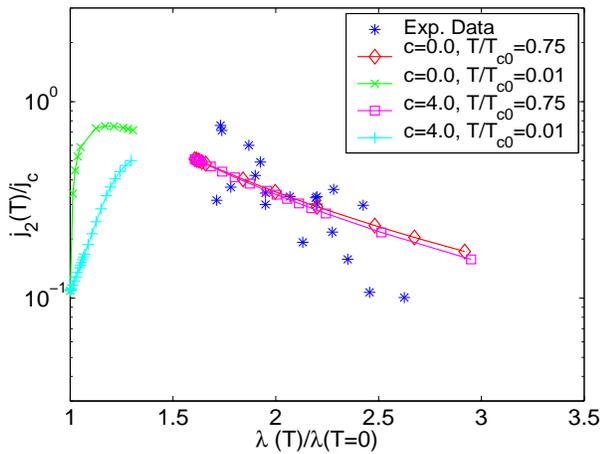}
\caption{(Color online) Comparison of the experimental points
shown in Fig. \ref{j2vslambda} and the theoretical results for the
intrinsic nonlinear current density scale $j_2$ versus the
(zero-current) penetration depth $\lambda(T)/\lambda(T=0)$. The
four curves correspond to unitary ($c=0.0$) and Born ($c=4.0$)
scatterers in the low ($+,\times$) and high ($\Diamond$,$\square$)
temperature limit.}\label{j2vslambdateo}
\end{figure}
In order to model the varying crystal quality of the samples used
to obtain the results shown in Fig. \ref{j2vslambda} we vary the
impurity parameters $\Gamma$, $c$ and use Eq.
(\ref{lambdadef})-(\ref{betadef}) to extract the penetration depth
$\lambda$ (at zero current) and the nonlinear current density
scale $j_2$. In Fig. \ref{j2vslambdateo} we show $j_2/j_c$ at
fixed temperature $T$ versus $\lambda(T)/\lambda(T=0)$ obtained
using this procedure. We show the results both for unitary
scatterers and in the Born limit. We have superimposed the
experimental data points from Fig. \ref{j2vslambda} by scaling
them with $j_c=150 \mbox{MA/cm$^2$}$ and $\lambda(T=0)=150
\mbox{nm}$ similar to the values given in the
literature\cite{DahmScalaJAP,Claassen2,hammond,mgb2}. As seen in
Fig. \ref{j2vslambdateo}, the high temperature results agree
reasonably well with the experiments. At high temperatures there
is little difference between unitary and Born scatterers, as
expected. For the dirtiest samples one can speculate whether
effects other than the intrinsic nonlinearity become important.
Certainly, the data point at $\lambda=480 \mbox{nm}$ (see Fig.
\ref{j2vslambda}) seems to fall outside the fit, which may be
related to the fact that this film was grown on sapphire.

\begin{figure}[t]
\begin{center}
\leavevmode
\begin{minipage}{.49\columnwidth}
\includegraphics[clip=true,width=.99\columnwidth]{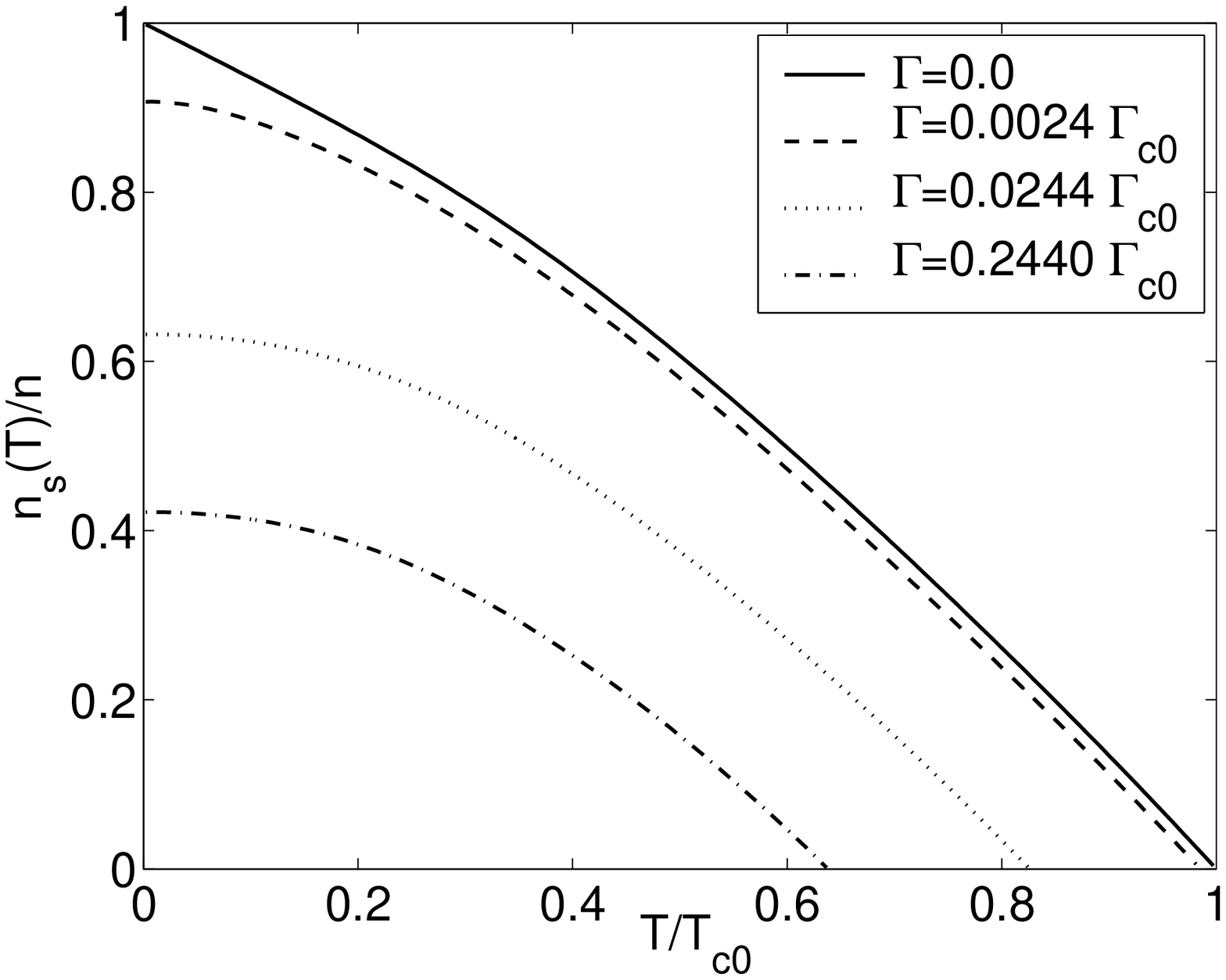}
\end{minipage}
\begin{minipage}{.49\columnwidth}
\includegraphics[clip=true,width=.99\columnwidth]{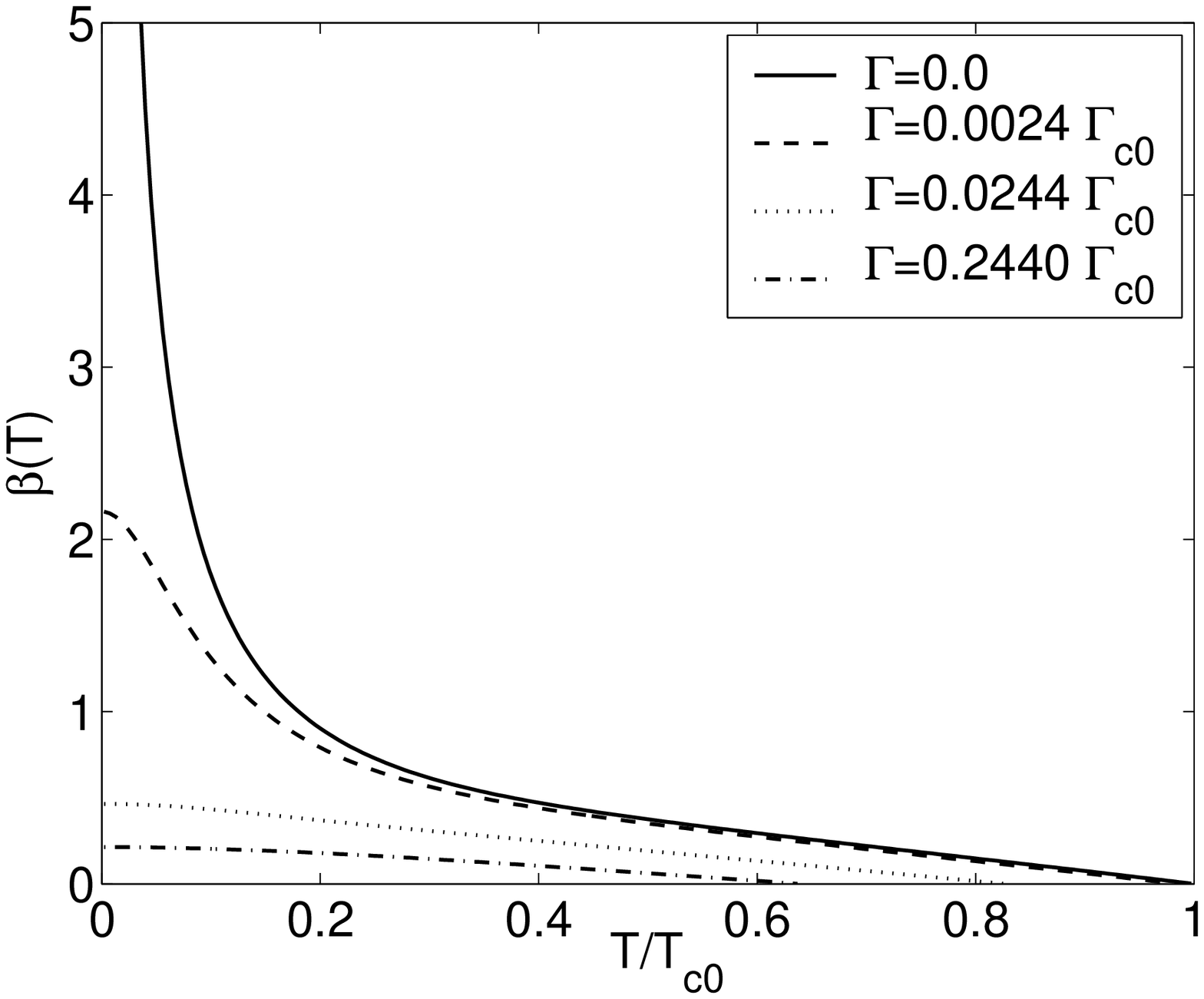}
\end{minipage}
\caption{$n_s(T)$ and $\beta(T)$ versus temperature $T$ for a
range of scattering strengths $\Gamma$ in the case $c=0$. Here,
$T$ and $\Gamma$ are given in units of the critical temperature of
the clean system $T_{c0}$ and the critical scattering strength
$\Gamma_{c0}$ at zero temperature, respectively. The right panel
is similar to Fig. 6 of Dahm and
Scalapino\cite{DahmScalaPRB}.}\label{nsbetac0theo}
\end{center}
\end{figure}

The nonlinear current density scale $j_2$ has a dramatically
different dependence on $\lambda$ at low temperatures $T \ll
T_{c0}$, where $T_{c0}$ is the critical temperature of the clean
system. Provided $\mbox{max} (\gamma,T) \gg \Delta_0 j_s/j_c$,
where $\gamma=-\mbox{Im} \tilde\omega(\omega=0)$ is the residual
quasiparticle width, we can discuss this behavior within the
quadratic approximation of Eq.(\ref{lambdaexp}). The functional
form of the low $T$ results shown in Fig. \ref{j2vslambdateo} can
be understood in terms of Eq.(\ref{j2def}) from Fig.
\ref{nsbetac0theo}, where we show the temperature dependence of
the superfluid density $n_s(T)$ and $\beta(T)$ in the unitary
limit for various $\Gamma$. In the clean limit $\Gamma=0.0$, the
$\beta(T)$ coefficient diverges at low temperatures due to the
nodes of the d-wave gap\cite{DahmScalaPRB}. With increasing
disorder, this divergence is strongly suppressed. At higher
temperatures, the $\Gamma$ dependence of $\beta(T)$ becomes much
weaker. On the other hand, the superfluid density $n_s$ in Fig.
\ref{nsbetac0theo} exhibits a nonsingular dependence on $\Gamma$.
From Eq.(\ref{j2def}) it is clear that the low $T$ divergence of
$\beta(T)$ eventually will drive $j_2$ down. This yields a
prediction for low $T \ll T_{c0}$ measurements valid if the
samples with lowest penetration depth in Fig. \ref{j2vslambda} are
clean enough to exhibit a pronounced $\beta(T)$ upturn at low
temperatures. It is clear from Fig. \ref{j2vslambdateo} that low
temperature measurements will provide a more stringent test of the
intrinsic nature of the nonlinear response in these materials, and
allow one to extract information on the nature of the defects
limiting the transport.

In summary, we have presented experimental results for the
nonlinear current density scale $j_2$ extracted from a range of
different quality YBCO samples. We have argued that $j_2$ is
consistent with an intrinsic origin. A calculation including
disorder within the self-consistent T-matrix approximation
provides reasonable agreement with these experiments.

{\it Acknowledgments.} We thank D. J. Scalapino for valuable
discussions. This work is supported by ONR grant N00014-04-0060.
This work is a contribution of the U.S. government, not subject to
copyright.

\end{document}